\documentclass[10pt]{iopart}
\usepackage{graphicx,iopams,bm,color,setstack,mathptmx}
\usepackage{soul}

\newcommand{\openone}{\leavevmode\hbox{\small1\normalsize\kern-.33em1}}

\newcommand{\matriz}[1]{\mathsf{#1}}

\begin{document}

\title{When quantum state tomography benefits from willful ignorance}

\author{Libor~Motka$^{1}$, Martin~Pa\'ur$^{1}$, Jaroslav~\v{R}eh\'{a}\v{c}ek$^{1}$, Zden\v{e}k~Hradil$^{1}$ and Luis~L~S\'{a}nchez-Soto$^{2,3}$}

\address{$^{1}$ Department of Optics, Palacky University, 17. listopadu 12, 77146 Olomouc, Czech Republic}
\address{$^{2}$ Departamento de \'Optica, Facultad de F\'{\i}sica, Universidad Complutense, 28040~Madrid, Spain}
\address{$^{3}$ Max-Planck-Institut f\"ur die Physik des Lichts,  91058~Erlangen, Germany}

\ead{rehacek@optics.upol.cz}
\vspace{10pt}
\begin{indented}
\item[]\today
\end{indented}

\begin{abstract}
We show that quantum state tomography with perfect knowledge of the measurement apparatus proves to be, in some instances, inferior to strategies discarding all information about the measurement at hand, as in the case of data pattern tomography. In those scenarios, the larger uncertainty about the measurement is traded for the smaller uncertainty about the reconstructed signal. This effect is more pronounced for minimal or nearly minimal informationally complete measurement settings, which are of utmost practical importance.
\end{abstract}
\eqnobysec
\vspace{2pc}

\submitto{New Journal of Physics}

\section{Introduction}

The goal of quantum state tomography (QST) is to infer a reliable estimate of the state of a quantum system from a suitable set of measurements performed on a finite set of identical copies of the system~\cite{lnp:2004uq,Teo:2015aa}. This technique has evolved from the first theoretical and experimental concepts to a fairly standard method~\cite{Lvovsky:2009ys,Toninelli:2019aa}.

The rapid progress in quantum-enhanced technologies entails the use of increasingly complicated systems, which in turn demands ever more sophisticated measurements. Although efficient procedures for QST are available, a persistent problem faced by contemporary QST is the complete characterization of the measurement. Having efficient and precise quantum detector calibration methods is particularly challenging: this is precisely the objective of quantum detector tomography (QDT)~\cite{Luis:1999yg,Fiurasek:2001dn,DAriano:2004oe}. 

The standard QDT reconstructs the action of the measurement from the outcome statistics in response to a set of complete certified input probes~\cite{Lundeen:2009sf,Zhang:2012fu}. This has been successfully applied to a variety of situations~\cite{Feito:2009aa,Coldenstrodt:2009aa,Akhlaghi:2011aa,DAuria:2011aa,Zhang:2012aa,Renema:2012aa,Natarajan:2013bh,Renema:2014aa}. However, QDT soon becomes exceedingly onerous and impractical as the number of detector outcomes grows. 

This drawback can be partially overcome with a more sophisticated calibration embracing the utilization of intrinsic quantum resources, such as entanglement. Examples include the absolute calibration method based on twin beams~\cite{Klyshko:1980aa,Worsley:2009aa,Avella:2011aa} and self-testing or blind tomography~\cite{Yang:2014aa,Wu:2016aa,Chen:2016aa}. Trading knowledge of probes for information about the measurement yields the concept of self-calibrating tomography~\cite{Mogilevtsev:2009aa,Mogilevtsev:2010aa,Branczyk:2012aa,Mogilevtsev:2012aa,Stark:2016aa,Sim:2019aa}.

When the measurement itself is of no interest, the expensive detector calibration can be overcome by using a direct fitting of the measure responses (patterns) to a set of known quantum states.  This is the essence of the data pattern tomography (DPT)~\cite{Rehacek:2010fk,Mogilevtsev:2013kl,Motka:2014aa,Motka:2017aa,Reut:2017aa}, which has been implemented with remarkable success~\cite{Cooper:2014aa,Harder:2014aa,Altorio:2016aa}. In this way, QST is accomplished without any prior knowledge of the measurement, avoiding unnecessary wasting of resources. Moreover, this approach is insensitive to the imperfections of the setup, because they are automatically accounted for by the  data patterns.

A quantitative comparison of DPT and QDT+QST strategies was carried out  in \cite{Motka:2017aa}, where it was demonstrated that DPT overcame  QDT+QST in many regimes of practical importance. Here, we take the case further by showing that DPT can even outperform QST suplemented with perfect knowledge about the measurement apparatus. This is a bit counterintuitive, as one would think that a complete description of the measurement should give a substantial advantage over an approximation spoiled by the measurement noise. Showing this is not universally true is the main result of the present paper.

{We proceed by analyzing the distinctions between QDT and DPT and by  giving an intuitive explanation of the above-mentioned  unexpected behavior. The rationale lies in the subtle effects of the omnipresent noise:} the noisy data corresponding to probes allows for a better fitting of the noisy data of an unknown state than the exact knowledge of the detection process.  In this sense, the \emph{exact} information about detection is just an apparent  advantage, which is causing a bias in the reconstruction scheme. For simplicity reasons, we adopt a linear inversion (being aware that it is not the optimal strategy), for it allows for .a deeper comprehension of the physics involved. We confirm our theoretical findings with numerical simulations of realistic setups.

\section{Quantum state tomography with unknown detectors}

Let us first set the stage for our analysis. {We deal with an input signal described by a true density matrix $\varrho$. Considering a $d$-dimensional quantum system, the positive semidefinite $d \times d$ matrix $\varrho$ requires $n \equiv d^{2} - 1$ independent real numbers for its specification. The following analysis can also be applied to continuous-variable systems provided the state space can be truncated in a suitable computational basis, so that all significant state components are contained within that finite-dimensional representation.}

QST aims at estimating $\varrho$ from measurements performed on identically prepared copies of the system under certification. In general, these measurements are represented by positive operator-valued measures (POVMs)~\cite{Helstrom:1976ij}. They are a set of Hermitian operators $\{ \Pi_j \}$ (with $\Pi_j\ge 0$ and $\sum_j \Pi_j= \openone$), such that each POVM element represents a single output channel of the measuring apparatus. We take every measurement as yielding $m$ distinct outcomes, the probability of detecting the $j$th output being given by Born's rule $p_{j} = \Tr (\varrho \Pi_{j})$.  

We can expand both $\varrho$ and the POVM $\{ \Pi_j \}$ in a suitable operator basis. A very convenient choice is a traceless Hermitian basis $\{\Gamma_{k}\}$ ($k=1, \ldots, n$), satisfying $\Tr (\Gamma_{k})= 0$ and $\Tr (\Gamma_{k} \Gamma_{\ell}) = \delta_{k\ell}$. This set coincides with the orthogonal generators of SU($d$), which is the associated symmetry algebra. In this way, we directly get
\begin{equation} 
\label{eq:tomo}
\bi{p} = \matriz{A} \, \bi{r},
\end{equation}
where we have omitted a trivial constant and $r_{k} = \Tr (\varrho \Gamma_k)$ and $\matriz{A}_{k\ell} = \Tr ( \Pi_{k} \Gamma_{\ell})$. $\matriz{A} $ is a unique $ m \times n$ real matrix that describes the explicit relation between the {theoretical probabilities $\bi{p}$} and the state parameters $\bi{r}$. 

In presence of noise and with a finite number of copies, {the collected relative frequencies of individual measurement outcomes},
we call $\bi{f}$, deviates from {their} expected values $\bi{p}$. The ultimate goal of tomography is to infer the unknown signal parameters $\bi{r}$ from the measured noisy data $\bi{f}$.  A number of different techniques are accessible, all of them providing a sensible inversion of equation~(\ref{eq:tomo}).  To facilitate as much as possible the analysis while retaining the fundamental features of the problem, we adopt here linear inversion: this can be accomplished by using either the ordinary least-square estimator (OLS)  or the generalized least-square estimator (GLS)~\cite{Lawson:1974aa}:
\begin{equation}
\widehat{\bi{r}}_{\mathrm{OLS}} = \matrix{A}^{+} \, \bi{f}  \, ,
\qquad \qquad
\widehat{\bi{r}}_{\mathrm{GLS}} = (\matriz{C}^{-1} \matriz{A})^{+} 
\matriz{C}^{-1} \bi{f} \, .
\end{equation} 
Henceforth, the hat will mean estimator,  $\matriz{G}= \matriz{C} \matriz{C}^{\dagger}$ is the data covariance matrix and the superscripts $\dagger$ and $+$ denote the Hermitian conjugation and the Moore-Penrose pseudoinverse~\cite{Penrose:1955aa,Ben-Israel:1977aa,Campbell:1991aa}, respectively. We just mention that the GLS is the best linear unbiased estimator {under the assumption of zero-mean noise $\langle f\rangle=p$ (with $\langle \cdot \rangle$ denoting average over data), we adopt throughout this paper, whereas OLS is a handy estimator for small and medium-sized data sets, when a reliable estimation of the data covariances is not possible~\cite{Kay:1993aa}.}

It is important to underline that the positivity of the reported density matrix $\widehat{\varrho}$ is not guaranteed by linear inversion methods. As positivity constraints serve as a kind of regularization, this leads to a slightly worse performance of OLS and GLS techniques on rank-deficient states compared to more elaborated statistically motivated inversion, such as maximum likelihood~\cite{Millar:2011aa} or Bayesian methods~\cite{Gelman:2013aa}. However, linear methods are enough for our purposes.  
{First, our preliminary analysis indicates that the effect discussed here persists for any positivity-constrained estimation. Second, differences between constrained and unconstrained estimators quickly disappear with the growing size of data measured on realistic (i.e., full rank) quantum states.}

Assume next that the QST is carried out with an unknown measurement apparatus. This may happen because either the details of the measurement are not known, or we choose to discard such information.  Two conceptually different ways of dealing with this task are at hand. In both of them, a set of $M$ known probes $\bi{r}^{(\alpha)}$ ($\alpha= 1, \ldots, M$) is measured and the corresponding data $\bi{f}^{(\alpha)}$, called \textit{patterns}, collected.  Arranging these probes and patterns columnwise, we get the $n\times M$ probe matrix $\matriz{R}$ and the $m\times M$ pattern matrix $\matriz{F}$. 

In what follows, we are interested in informationally complete schemes, so that any quantum state $\varrho$ can be unambiguously assigned to the corresponding theoretical probabilities $\bi{p}$. This requires the matrix $\matriz{A}$ to have at least $d^{2} - 1$ linearly independent rows, which means that $m, M\ge n$. In particular, the minimal or nearly minimal tomography  $n \approx m \ll M$ is of special interest, because  having a small number of measurement outputs improves the feasibility of the scheme. Highly overcomplete schemes may have a better performance, but their practical implementation can become a formidable task. 

The two protocols above mentioned are as follows:

(a) \textit{Detector-tomography-assisted quantum state tomography} (DQST). In DQST, we first implement QDT to estimate the measurement matrix from measured probes.  Therefore, we have to solve $\matriz{F} = \matriz{A}  \, \matriz{R}$.  The detector is thus specified by 
\begin{equation}
\matriz{A}_{s} = \matriz{F} \matriz{R}^{+} \, .
\end{equation} 
where we have attached the subscript $s$ to stress that we are working in this standard tomographic procedure. The next step is the QST, which is tantamount to solve (\ref{eq:tomo}) for the signal $\bi{r}$. If we use OLS (the generalization to GLS is straightforward), we get 
\begin{equation}
\label{eq:rs}
\widehat{\bi{r}}_s= \widehat{\matriz{A}}_{s}^{+} \bi{f} = 
(\matriz{F} \matriz{R}^{+})^{+} \bi{f} \, .
\end{equation}

(b) \textit{Data pattern tomography} (DPT). With the same set of probes and patterns, in DPT we bypass the QDT  and construct a best fit of the data $\bi{f}$ in terms of patterns $\matriz{F}$
{
and fitting coefficients $\bi{x}$
}
; i.e., $ \matriz{F} \bi{x} \simeq  \bi{f}$, which results in 
\begin{equation}
\label{patfit}
\widehat{\bi{x}} = \matriz{F}^{+}\, \bi{f} \, .
\end{equation}
Notice that when there are less measurement channels than probes, $m<M$, the set (\ref{patfit}) is underdetermined, and hence the OLS and GLS estimators of $\bi{x}$ coincide. The final state estimate is formed by combining the probes
\begin{equation}
\label{eq:rp}
\widehat{r}_p= \matriz{R} \, \hat{x} = \matriz{R} \matriz{F}^{+} \, \bi{f} \, .
\end{equation}
Here, we use the subscript $p$ to denote the DPT results. From (\ref{patfit}) it is clear that the effective measurement matrix in this method reads 
\begin{equation}
\matriz{A}_{p} = (\matriz{R} \matriz{F}^{+})^{+} \, .
\end{equation}
We stress once more that we are presenting a linear version of DPT. In real applications, the data fitting is subject to $\widehat{\varrho}\ge 0$ to produce a physically meaningful estimate. 

A glance at equations~(\ref{eq:rs}) and (\ref{eq:rp}) immediately shows that both methods become equivalent whenever 
\begin{equation}
  \label{equiv}
  ( \matriz{F} \matriz{R}^{+})^{+}  = \matriz{R} \matriz{F}^{+}  \, .
\end{equation}
This always holds true for regular matrix inverses, but not always by pseudoinverses. As demonstrated in \cite{Motka:2017aa}, the identity~(\ref{equiv}) holds true if $\matriz{F}$ and $\matriz{R}$ have full column rank.  

It is interesting to consider the behavior of both  DQST and DPT in the limit of a large number of probes $ M \rightarrow \infty$. Assuming an additive noise model for patterns; i.e., $\matriz{F}= \matriz{A} \matriz{R} + \matriz{S}$, with $\matriz{S}$ being the noise matrix,  the detector step of DQST becomes
\begin{equation}
\label{thelimit}
\lim_{M \rightarrow \infty} \widehat{\matriz{A}}_{s} = \matriz{A} + 
\lim_{M\rightarrow\infty} \matriz{S} \matriz{R}^{+} = \matriz{A}
\end{equation}  
for any noise strength. {On adding more patterns to $\matriz{F}$ (i.e., increasing the number of columns $M$ while preserving the number of rows $m$) makes all singular values of $\matriz{F}$ and $\matriz{F}^{+}$ behave like $O(\sqrt{M})$ and $O(1/\sqrt{M})$, respectively. A random walk-like process introduced by multiplying $\matriz{F}^{+}$ with the noise matrix $\matriz{S}$ results in the trace-class norm $||\matriz{S} \matriz{R}^{+}||$ going with $O(1/\sqrt{M})$, hence~(\ref{thelimit}).} So, by increasing the number of probes, the QDT converges to the true design  matrix $\matriz{A}$ and DQST becomes equivalent to QST with a perfect knowledge about the measurement apparatus: {this does not come as as surprise.}

{Taking the same limit of the effective inverse detector matrix $\widehat{A}^{+}_{p}= \matriz{R} \matriz{F}^{+}$ of DPT gives 
\begin{equation}
\fl
\label{thelimit2}
\lim_{M\rightarrow\infty}\widehat{A}^{+}_{p} = 
\lim_{M\rightarrow\infty} \matriz{R}( \matriz{A} \matriz{R}+ \matriz{S})^{+} \neq 
\lim_{M\rightarrow\infty} \left[(\matriz{A} \matriz{R}+\matriz{S})\matriz{R}^{+}\right]^{+}=
\lim_{M\rightarrow\infty} (\matriz{A} +\matriz{S}\matriz{R}^{+})^{+}=
\matriz{A}^{+} \,,
\end{equation}
where the inequality is due to the violation of (\ref{equiv}). The DPT inversion $\widehat{A}^{+}_{p}$, in general differs (is biased) from the true inverse $\matriz{A}^{+}$.}

What is crucial for us is that introducing such a bias can be beneficial and improve the performance of DPT with respect to DQST. Notice that the effective inverse 
$\widehat{\matriz{A}}^{+}_{p}$ is the OLS solution to the problem
\begin{equation}
\label{OLSproblem}
 \matriz{A}^{+}_{p} \matriz{F}= \matriz{R} \, .
\end{equation}
By the properties of OLS~\cite{Lawson:1974aa}, the DPT inverse provides the best reconstruction of probes from the measured patterns. Hence, in the limit of a large $M$, where the distance of any unknown state from the set of probes can be made arbitrarily small, the DPT becomes optimal. As will be shown below with numerical simulations, moderate values of $M$ are sufficient to see a significant advantage of the DPT over the DQST for nearly minimal informationally complete settings.

\section{Resolution limits}

The time-honored Cram\'{e}r-Rao lower bound (CRLB)~\cite{Rao:1945aa,Cramer:1946aa} can be used to bound the mean square error $e^{2}$ of any unbiased estimator $\widehat{\varrho}$ of the true density matrix $\varrho$:
\begin{equation}
 || \varrho-\widehat{\varrho} ||^{2} = ||\bi{r} - \widehat{\bi{r}} ||^2 
 \ge \Tr \left ( \mathcal{F}^{-1} \right ) = e^2_{\mathrm{CRLB}},
\end{equation}   
where $\mathcal{F}$ is the (classical) Fisher information matrix~\cite{Fisher:1925aa} associated with a given measurement and true state
\begin{equation}
\mathcal{F}_{kl}= \left\langle 
\left(\frac{\partial}{\partial r_{k}} \ln \mathcal{L}\right) 
\left(\frac{\partial}{\partial r_{l}} \ln\mathcal{L}\right)
\right\rangle \, .
\end{equation}
Here, $\mathcal{L}$ is the likelihood; i.e., the probability of registering data $\bi{f}$ given the true state $\bi{r}$.
For example, for Poissonian statistics with  $N$ events, each one with probability $p_{j}$, we have
\begin{equation}
\mathcal{L}=\prod_{j} \frac{(N p_j)^{N_j}}{N_j!} e^{-N p_j} \, ,
\end{equation}
and hence
\begin{equation}
\mathcal{F} = N \matriz{A}^{\dagger} \bar{\matriz{P}} \matriz{A}, \qquad \qquad
\bar{\matriz{P}}_{jj^{\prime}} = \delta_{jj^{\prime}} p_{j}^{-1}.
\end{equation}
In the limit of many probes $M\rightarrow \infty$, the QDT step of DQST converges towards the true measurement matrix and so the DQST is expected to attain the bound of QST with perfect knowledge of the measurement apparatus. {However, in this case the DPT estimator becomes biased and the CRLB must be suitably modified~\cite{Eldar:2004vj}.  There are many instances where a biased estimator displays better performance than the predicted by the  unbiased CRLB~\cite{Stoica:1990uw,MacEachern:1993vj,Motka:2016aa}. A particularly relevant one is the Rayleigh curse in estimating the separation between two incoherent pointlike sources~\cite{Tsang:2016aa,Tsang:2018aa}.}

%%%%%%%%%%%%%%%%%%%%%%%%%%%%%%%%%%%%%%%
\begin{figure}[t]
\centerline{\includegraphics[width=0.47\columnwidth]{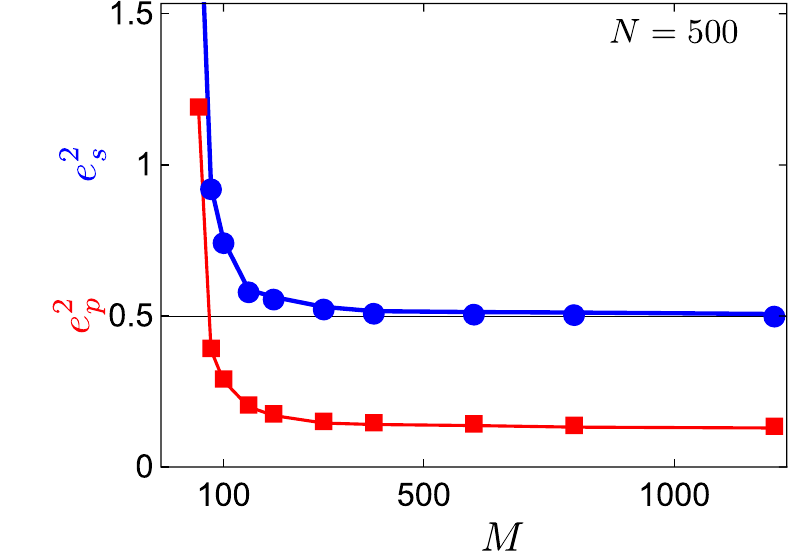} 
\includegraphics[width=0.47\columnwidth]{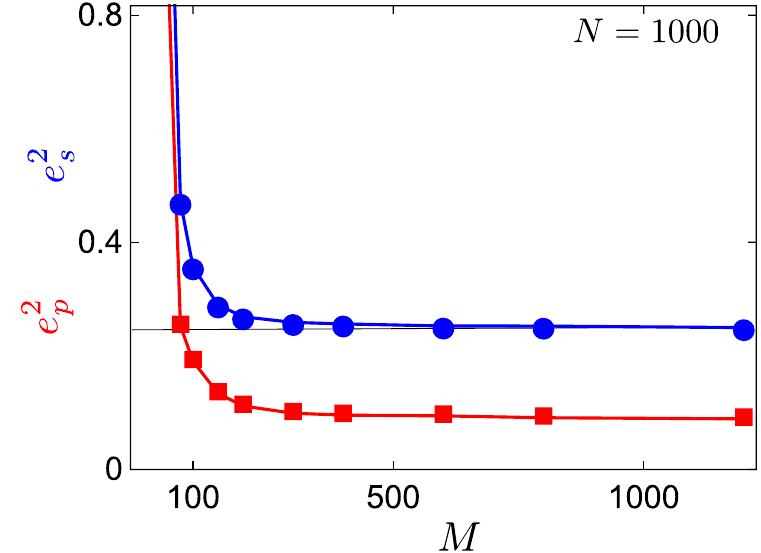}}
\centerline{\includegraphics[width=0.47\columnwidth]{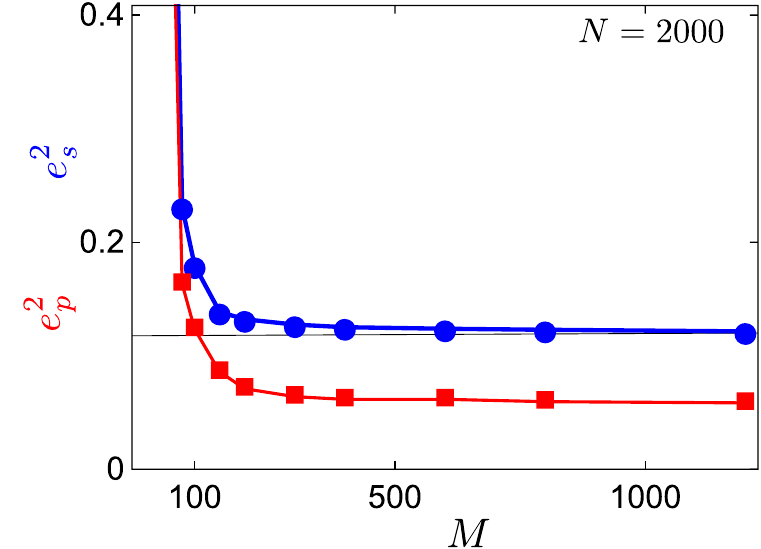}
\includegraphics[width=0.47\columnwidth]{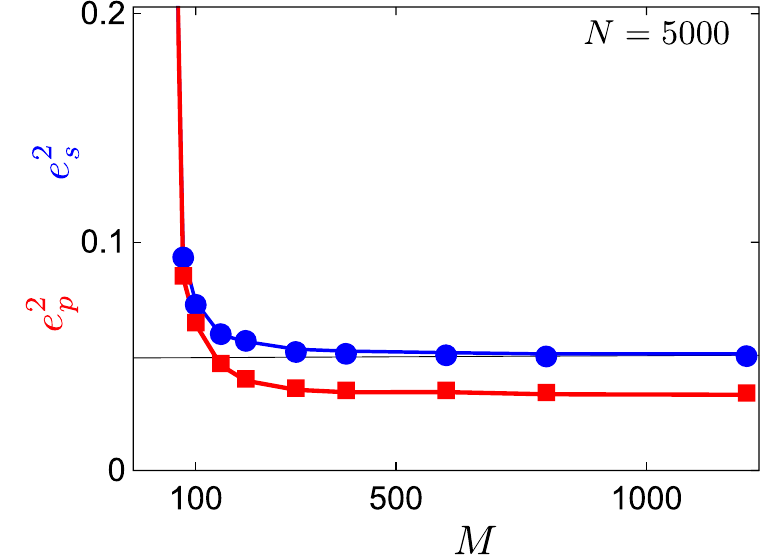}}
\caption{Mean square errors of the DQST (blue circles) and DPT (red squares) methods versus the size of the probe set. The black horizontal line is the CRLB of the QST with a perfectly-known apparatus. The total number of detection events is indicated in the corresponding insets. See the text for more details.} 
\label{fig:results}
\end{figure}
%%%%%%%%%%%%%%%%%%%%%%%%%%%%%%%%%%%%
%%%%%%%%%%%%%%%%%%%%%%%%%%%
\begin{figure}[b]
\centerline{\includegraphics[width=0.65\columnwidth]{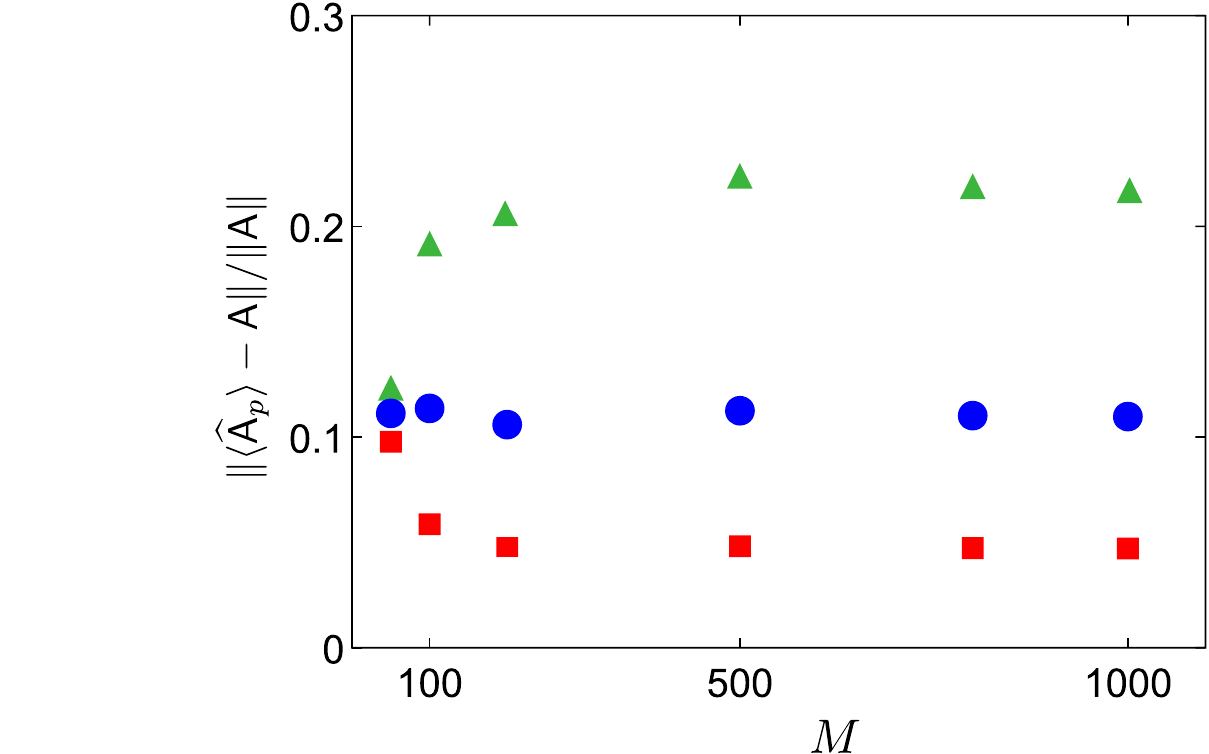}}
\caption{Bias of the DPT estimators of the design matrix $\matriz{A}$ versus the size of the probe set for $N=500$ (green triangles), $1000$ (blue circles), and $3000$ (red squares) detection events. The same random square root measurement as in figure~\ref{fig:results} was used.}
\label{fig:bias}
\end{figure}
%%%%%%%%%%%%%%%%%%%%%%%%%%

\section{Examples and discussion}

Excellent performance of the DPT for nearly minimal informationally complete measurements can be illustrated with numerical experiments. We use random square-root measurements defined as 
\begin{equation}
\Pi_{j} = G^{-1/2} |\phi_{j} \rangle \langle \phi_{j}| G^{-1/2} \,, 
\qquad \qquad
G = \sum_{j}|\phi_{j} \rangle \langle \phi_{j}| \, ,
\end{equation}
where $|\phi_{j} \rangle$ are randomly generated Haar distributed pure states and $j= 1, \ldots, m$.  This has been proposed as  \emph{a pretty good} measurement for distinguishing quantum states~\cite{Hausladen:1994aa} and is known to be optimal~\cite{Dalla-Pozza:2015aa}. In our case, we take $m=40$ outputs applied to a six-dimensional quantum system prepared in one random pure state with a $10\%$ admixture of the maximally mixed state to make it full rank.  Notice that the measurement is nearly minimal $m \approx 6^2$. We assume Poissonian detection statistics with $N$ total detection events and calculate the mean square error of the DPT and DQST estimates from $1000$ repeated data acquisitions. We also calculate the CRLB for the QST with perfect knowledge of the measurement apparatus.

The results are summarized in figure~\ref{fig:results} for $N=$ 500, 1000, 2000 and 5000 events. The advantage of the DPT over DQST for moderate and large numbers of probes is evident. The corresponding fidelity pairs $[\mathrm{fidDQST, fidDPT}]$ for the large $M$ limit  are $[86.2\%,90.7\%]$, $[89.7\%,92.5\%]$, $[92.5\%,94.1\%]$, and $[94.9\%,95.5\%]$ and hence to match the fidelity of the DPT with the standard DQST technique, about twice as many detections must be registered. 

{Notice also that} for moderate and large sets of probes, the DPT errors are squeezed below the CRLB limit.  {This means that instead of using perfect information about the measurement apparatus for QST, the better strategy is to discard that information, probe the measurement device anew with a sufficiently large probe set and use the DPT approach. In fact, since the CRLB applies to any unbiased estimator, any unbiased QST based on the true design matrix is outperformed by DPT in that regime. 
}

The observed CRLB violation can be explained by the bias inherent to the DPT data processing. This is illustrated in figure~\ref{fig:bias}, where the bias of the effective DPT measurement estimator $\widehat{\matriz{A}}_p = (\matriz{R} \matriz{F}^{+})^{+}$ is shown for different values of $M$ and $N$. The bias tends to zero with increasing the sample size. However, for any fixed noise strength, the bias is nonzero and remains so even for large sets of probes. This bias, in turn, makes the DPT state estimator biased and  the standard CRLB does not apply. The apparently \emph{wrong} measurement estimate is of no concern in DPT as we are not interested in the description of the measurement apparatus but only in the final state estimate.

%%%%%%%%%%%%%%%%%%%%%%%%%%%%%
\begin{figure}[t]
\centerline{\includegraphics[width=0.56\columnwidth]{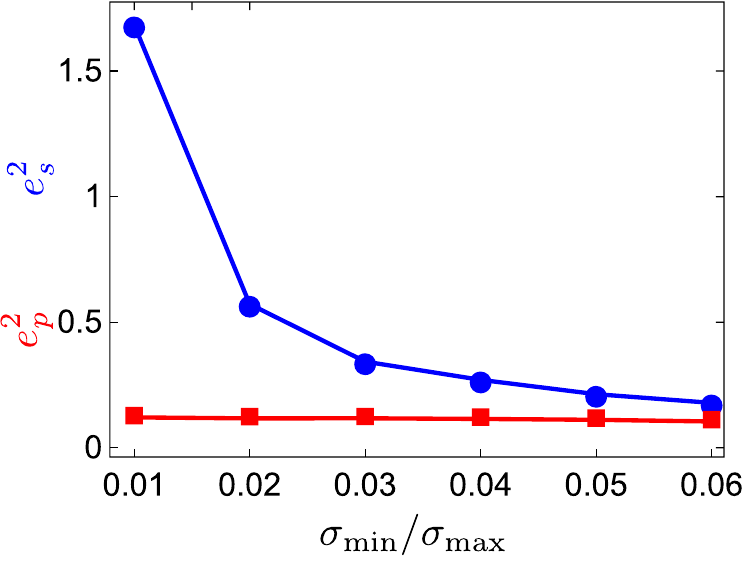}}
\caption{Mean square errors of the DQST (blue circles) and DPT (red squares) vs the inverse condition number of the measurement design matrix $\matriz{A}$. Each point is obtained by averaging over $1000$ randomly generated square-root measurements with inverse condition numbers within $0.001$ of the indicated value, $N=1000$ detection events per measurement and $M = 1200$ probes. Other relevant parameters are like those used for generating figure~\ref{fig:results}. For the large number of probes used here, the blue symbols are very close to their respective CRLB bounds and so the latter are omitted for clarity. \label{fig:illposed}}
\end{figure}
%%%%%%%%%%%%%%%%%%%%%%

Having seen the performance gap between the DPT and DQST growing with the noise strength, we note that a similar effect can be observed for badly conditioned measurements, as can be appreciated in figure~\ref{fig:illposed}. Here, the noise strength is fixed and we take a large set of probes to simulate the asymptotic limit. Increasing the condition number of the true design matrix makes the linear system (\ref{eq:tomo}) ill-posed, and this is reflected in the CRLB and, as a consequence, in the rapid growth of estimation errors of the standard DQST technique. The alternative DPT approach is much less sensitive to measurement quality and due to its biased nature can produce errors squeezed much below those of DQST but also much below the CRLB bounds derived for perfect knowledge about the measurement apparatus. 

On account of the figures~\ref{fig:results} and \ref{fig:illposed}, we arrive at a counterintuitive conclusion: to get a better QST in some scenarios, one is advised to discard the true description of the measurement apparatus and replace it with a crude and biased approximation. The resolution of this apparent contradiction is as follows: in the DPT approach, the noisy data measured is matched to noisy data patterns. In this way, the estimation technique is trained by noisy patterns to handle the noisy data better than if we used the true (correct) description of the measurement apparatus. The latter being equivalent to matching the noisy data to noiseless patterns. Interestingly, training the estimation procedure to cope with the noise in an optimal way seems equivalent to using a wrong (biased) description of the measurement device for QST. Errors introduced in an intermediate stage improve the final result!

\section{Concluding remarks}

In summary, we have demonstrated the remarkable performance of DPT. Probing an unknown measurement device with a sufficiently large set of probes in some regimes of practical interest turns out to be a better strategy than adopting a perfect knowledge about the measurement apparatus. In other words, discarding true information at hand and replacing it with a crude approximation leads to better final results. This is possible because of the way the DPT self-adapts to the omnipresent noise. We confirmed our findings with numerical simulations of random square-root measurements. 
{Similar behavior is expected for other kinds of measurements and noise.}

\ack

This work was supported by the European Union's Horizon 2020 research and innovation programme under the QuantERA programme through the project ApresSF and from the EU Grant  899587 (Project Stormytune), the Grant Agency of the Czech Republic (Grant 18-04291S), The Palacky University Grant {IGA}\_PrF\_2020\_004 and the Spanish Ministerio de Ciencia e Innovacion (Grant PGC2018-099183-B-I00).

\newpage

%\bibliographystyle{unsrt}
%\bibliography{Tomo}

\end{document}